\documentclass[a4paper,11pt]{article}

\advance\textheight by \topskip
\usepackage{rotating}

\begin{document}

\markboth{\bf Vortex deformation and breaking in superconductors: A
microscopic description}{E. Pardo, J. H. Durrell and M. G.
Blamire}

\title {Vortex deformation and breaking in superconductors: A microscopic description}

\author{E. Pardo$^{a,b}$, J. H. Durrell$^a$ and M. G. Blamire$^a$\\
$^a$Department of Materials Science and Metallurgy,\\ University of
Cambridge, Pembroke Street, Cambridge,\\CB2 3QZ, United Kingdom\\
$^b$Grup d'Electromagnetisme, Departament de F\'\i sica,\\ Universitat
Aut\`onoma de Barcelona, 08193 Bellaterra,\\ Barcelona, Catalonia,
Spain}

\maketitle

\begin{abstract}
Vortex breaking has been traditionally studied for nonuniform
critical current densities, although it may also appear due to
nonuniform pinning force distributions. In this article we study
the case of a high-pinning/low-pinning/high-pinning layered
structure. We have developed an elastic model for describing the
deformation of a vortex in these systems in the presence of a
uniform transport current density $J$ for any arbitrary
orientation of the transport current and the magnetic field. If
$J$ is above a certain critical value, $J_c$, the vortex breaks
and a finite effective resistance appears. Our model can be
applied to some experimental configurations where vortex breaking
naturally exists. This is the case for YBa$_2$Cu$_3$O$_{7-\delta}$
(YBCO) low angle grain boundaries and films on vicinal substrates,
where the breaking is experienced by Abrikosov-Josephson vortices
(AJV) and Josephson string vortices (SV), respectively. With our
model, we have experimentally extracted some intrinsic parameters
of the AJV and SV, such as the line tension $\epsilon_l$ and
compared it to existing predictions based on the vortex structure.
\end{abstract}

\tableofcontents

\section{Introduction}

The study of vortex physics has been an important topic in
superconductivity research since the prediction of the existance of
superconducting vortices by Abrikosov
\cite{abrikosov57JTP,campbell72APh,blatter94RMP,brandt95RPP}.
Superconducting vortices have been the subject of many theoretical
studies including those addressing the vortex line energy
\cite{clem75JLT,huCR72PRB}, single vortex interactions
\cite{brandt79JLT,clem80JLT,wagenleithner82JLT,boudiab01PRL} and
vortex lattice interactions
\cite{brandt77JLTa,brandt77JLTb,sudbo91PRL,sudbo91PRB}.

\begin{figure}[htb]
\begin{center}
\vspace{-0cm}
\includegraphics[width=12.5cm]{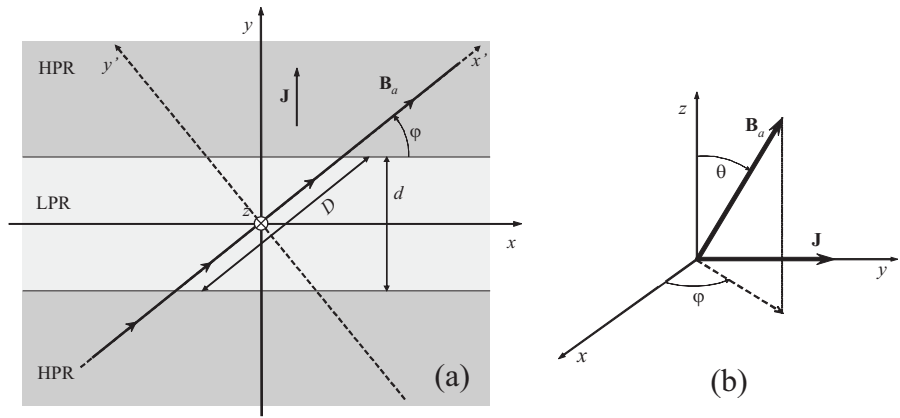}
\vspace{0cm} \caption{\small (a) Sketch of an undeformed vortex line
in a depressed pinning layer, consisting of a low pinning region
(LPR) bounded by high pinning ones (HPR). The applied field ${\bf
B}_a$ makes an angle with both the LPR-HPR interface and $\bf J$.
The vortex length in the LPR, $D$, is $D=d/|\sin\theta\sin\varphi|$.
(b) Definition of the angular coordinates $\varphi$ and $\theta$.}
\label{f.sketchGB}
\end{center}
\end{figure}

One way to measure the properties of a vortex, such as the vortex
line tension $\epsilon_l$ is by means of vortex breaking
experiments. These kind of experiments were initially performed by
using a nonuniform transport current in order to create a nonuniform
driving force in the vortex length thus bending the vortex and, for
large enough current density difference, break (or cut) the vortex
\cite{ekin75PRB,busch92PRL,lopez94PRB,blamire86PRB,grigera02PhC}.
Another way in which vortex breaking may be observed is where
vortices are subject to an inhomogenous pinning force in the
presence of a uniform transport current. In this article, we study
the microscopic bending and breaking process for the situation where
the vortex crosses a low-pinning region (LPR) of thickness $d$
compressed by two high-pinning ones (HPR) with a uniform transport
current $\bf J$ and a magnetic field $\bf B$ with an arbitrary
orientation, Fig. \ref{f.sketchGB}. In order to perform this study,
an elastic model is developed assuming isotropic vortices with a
constant line tension $\epsilon_l$ towards deformations. Using this
model, we propose a vortex breaking process in the depressed pinning
region. Afterwards, a $J_c(\theta,\varphi)$ dependence is obtained,
where $\theta$ and $\varphi$ are the angles defining the orientation
of the applied field (see Fig. \ref{f.sketchGB}), as a function of
the elastic force per unit length and the pinning line forces in
either the LPR and the HPR.

The geometry described in this article can be directly applied in
order to model experimental situations like low-angle grain
boundaries (LAGB) in YBa$_2$Cu$_3$O$_{7-\delta}$ (YBCO) bicrystals
or YBCO films on a miscut substrate (commonly named vicinal
films), as discussed in Sec. \ref{s.exp}. It is worth noting that,
according to Gurevich {\it et al} \cite{gurevich02PRL} the
vortices in high-temperature superconductor LAGB for crystal
misorientation angles up to 23${\rm ^o}$ are not conventional
Abrikosov vortices (AV) but Abrikosov-Josephson vortices
(AJV)\footnote{Actually, the structure of a vortex in a LAGB for
anisotropic superconductors is more complex than the AJV described
in \cite{gurevich02PRL}. The vortex nature in our studied
situation is discussed in Sec. \ref{ss.expGB}.}. Moreover, the
vortices that it is possible to break in YBCO vicinal films are
string vortices (SV) lying between the $ab$ planes, which are of
Josephson nature. In this paper, the vortex line tension
$\epsilon_l$ in these systems is found by fitting the
theoretically obtained $J_c(\theta,\varphi)$ dependence to
measured data. Then, thanks to the model developed in this work,
it is possible to extract the line tension of exotic
superconductors of Josephson and Abrikosov-Josephson nature, which
physical properties are still not very well known.

The article is structured as follows. First, some general features
of our problem, including the basis of the elastic model, are
discussed in Sec. \ref{s.gen}. In Sec. \ref{s.analit} we deduce the
analytical limits of the vortex deformation and critical current
density for the studied HPR/LPR/HPR structure at two specific
magnetic field orientations: $\bf B$ perpendicular or parallel to
$\bf J$, with $\bf J$ crossing the HPR/LPR boundary. In the
following section, Sec. \ref{s.num}, we introduce a numerical
procedure for solving the problem for any field and current
orientation and we present general normalized results. Next, Sec.
\ref{s.exp}, the model is applied to the experimental situations of
LAGB in YBCO bicrystals and vicinal films in order to obtain the AJV
and SV vortex line tension, respectively. In that section, it is
also discussed the flux breaking process in a vicinal film. Finally,
we present our conclusions in Sec. \ref{s.concl}.


\section{General considerations}
\label{s.gen}

In this section we discuss some general features of our system, such
as the basis of the elastic model and the properties of the vortex
line energy.


\subsection{Elastic model}
\label{ss.elmod}

We simulate the behaviour of a vortex line as an string with
constant line energy $\epsilon_l$. Thus, its total energy is
\begin{equation}
\label{Eel} E={\epsilon_l}\int_C{\rm d}l={\epsilon_l}\int_C{\rm
d}{\tau} |\partial_\tau{\bf r}|,
\end{equation}
where $C$ is the vortex line path, $l$ is the line length and
$\tau$ is a parametrization of $C$. If the vortex follows a
straight line, the elastic force at the vortex ends is
\begin{equation}
{\bf F}_e=\pm{\bf e}_l{{\rm d} E}/{{\rm d} l}=\pm{\epsilon_l}{\bf
e}_l,
\end{equation}
where ${\bf e}_l$ is the unit vector locally parallel to the
vortex line:
\begin{equation}
\label{el} {\bf e}_l(\tau)=\frac{\partial_\tau{\bf
r}}{|\partial_\tau{\bf r}|}.
\end{equation}
From Eq. (2) it is clear that for an isotropic string the line
tension equals to the line energy. The local elastic force per
unit length ${\bf f}_e(\tau)$ for a deformed vortex can be deduced
by approximating the vortex line into a set of straight segments
and using that the net elastic force at a junction between
segments is
\begin{equation}
\label{delF} \delta{\bf F}_e(\tau)=\epsilon_l[{\bf e}_l(\tau)-{\bf
e}_l(\tau-\delta \tau)],
\end{equation}
where $\delta \tau$ is a variation in $\tau$ and $\delta{\bf
F}_e(\tau)$ is the elastic force in the junction between the
vortex segments corresponding to $\tau$ and $\tau-\delta \tau$.
After making the limit of $\delta \tau\to 0$, the force per unit
length can be found as
\begin{equation}
\label{ftau} {\bf f}_e(\tau)=\frac{\epsilon_l}{|\partial_\tau{\bf
r}|^4}\partial_\tau{\bf r}\times(\partial_\tau^2{\bf
r}\times\partial_\tau{\bf r}).
\end{equation}

Apart from the elastic force, the vortex is submitted to a driving
force, ${\bf F}_d$, due to a macroscopic current density $\bf J$.
The line force density of this force, ${\bf f}_d$, is
\begin{equation}
\label{fd} {\bf f}_d=\Phi_0{\bf J}\times{\bf e}_l,
\end{equation}
where $\Phi_0$ is the flux quantum, that in SI is $\Phi_0=\pi\hbar
e$.

For superconductors with pinning we should also take into account
the pinning force per unit length ${\bf f}_p$, which has a certain
maximum magnitude $f_{p,m}$. This force opposes any local movement
of the vortex line, so that it will follow the direction of the
elastic and the driving forces. This description of the pinning
behaviour implicitly assumes that ${\bf f}_p$ is due to many weak
pinning centres uniformly distributed in the vortex length, rather
than few strong pinning sites. The latter scenario would result in a
more complicated bending behaviour.

The equilibrium vortex configuration will be that one which
results in a null net force per unit length at any point. Thus,
from Eqs. (\ref{ftau}) and (\ref{fd}) we find the differential
equation of the vortex line
\begin{equation}
\label{difeq} \frac{\epsilon_l}{|\partial_\tau{\bf
r}|^4}\partial_\tau{\bf r}\times(\partial_\tau^2{\bf
r}\times\partial_\tau{\bf r})+\Phi_0{\bf J}\times{\bf e}_l+{\bf
f}_p=0.
\end{equation}


\subsection{On the line tension}
\label{ss.epl}

We next discuss the dependence of the line tension $\epsilon_l$ on
the superconductor internal parameters for both an isotropic and
anisotropic Abrikosov vortex (AV).

For an isolated isotropic AV, the line tension corresponds to the
energy per unit length, which has been investigated by several
authors \cite{clem75JLT,blatter94RMP,huCR72PRB}. Following the
numerical calculations of Hu \cite{huCR72PRB}, $\epsilon_l$ in the
approximation of high Ginzburg-Landau parameter $\kappa$ is
\begin{equation}
\label{elhk} \epsilon_l=\epsilon_{l,0}(\ln\kappa+0.4968),
\end{equation}
where $\epsilon_{l,0}=(\Phi_0/\lambda)^2/(4\pi\mu_0)$ and $\lambda$
is the magnetic penetration depth. A more general, although maybe
less accurate, analytical expression for arbitrary $\kappa$ was
obtained by Clem \cite{clem75JLT} using a variational approach. That
expression, however, is still proportional to $\epsilon_{l,0}$.

A vortex in a flux line lattice has a different line tension. In
general, $\epsilon_l$ depends on the vortex separation, $a_0$, and
the characteristic deformation wavelength, $l_d$
\cite{blatter94RMP,brandt95RPP}. The qualitative behaviour is that
$\epsilon_l$ decreases with decreasing both $a_0$ and $l_d$.
However, for magnetic fields such that $a_0\ll \lambda$ and
$a_0\gg \xi$, the line tension is roughly constant (where $\xi$ is
the superconductor coherence length). In a simplified way, this
can be explained as follows: if $a_0\ll \lambda$ the magnetic
field and the supercurrent density is roughly uniform between the
vortices except close to the vortex core. Then, the energy
variation per unit length due to a deformation is just the core
energy per unit length and, consequently, it does not depend on
the magnetic field. Following \cite{boudiab01PRL}, $\epsilon_l$
for the London limit is $\epsilon_l=\epsilon_{l,0}\int_1^\infty
{\rm d}x
(e^{-x/\kappa})/{\sqrt{x^2-1}}+\epsilon_{l,c}=\epsilon_{l,0}K_0(1/\kappa)+\epsilon_{l,c}$,
where $\epsilon_{l,c}$ is the core energy per unit length and
$K_0$ is the modified Bessel function of order 0. Comparing with
Eq. (\ref{elhk}), we obtain that
\begin{equation}
\epsilon_{l,c}=\epsilon_{l,0} \left[ \ln\kappa+0.4968-K_0(1/\kappa)
\right]. \label{elc}
\end{equation}
For the limit of $\kappa\gg 1$, Eq. (\ref{elc}) becomes\footnote{The
approximation of $\kappa\gg 1$ produces an error smaller than a 1\%
for $\kappa\ge 20$.} $\epsilon_{l,c}=0.3809\epsilon_{l,0}$, so that
we obtain a line tension of the order of $\epsilon_{l,0}$.

In Sec. \ref{s.exp} we study the line tension in SVs in YBCO, for
which $\epsilon_l$ is essentially the same as an anisotropic
Abrikosov vortex \cite{blatter94RMP,brandt95RPP}. Following Ref.
\cite{blatter94RMP}, the line tension for an anisotropic vortex
parallel to the $ab$ planes with $\kappa\gg 1$ and \footnote{For
our experimentally studied situations of Sec. \ref{s.exp}, the
condition $l_d\ll \lambda_{ab}$ is satisfied for $\varphi>5^{\rm
o}$ and $\theta>0.5^{\rm o}$ for LAGB and vicinal films,
respectively.} $l_d\ll \lambda_{ab}$, where $\lambda_{ab}$ is the
penetration depth for the $ab$ planes, is approximately
\begin{eqnarray}
\epsilon_l^\perp & \approx & \Gamma\frac{\Phi_0^2}{4\pi\mu_0\lambda_{ab}^2}\label{elSVt}\\
\epsilon_l^\parallel & \approx &
\frac{1}{\Gamma}\frac{\Phi_0^2}{4\pi\mu_0\lambda_{ab}^2},
\label{elSVp}
\end{eqnarray}
where $\epsilon_l^\perp$ and $\epsilon_l^\parallel$ are for
deformations in the $c$ and $ab$ directions, respectively, and
$\Gamma$ is the material anisotropy factor
$\Gamma=\lambda_c/\lambda_{ab}>1$ (for YBCO, $\Gamma\approx 5$),
where $\lambda_c$ is the penetration depth in the $c$ axis.


\section{Analytical limits}
\label{s.analit}

In this section we analytically describe the limits of $\varphi\to
0$ with $\theta=\pi/2$ (${\bf B}$ perpendicular to ${\bf J}$) and
$\varphi\to\pi/2$ with $\theta=\pi/2$ (${\bf B}$ parallel to ${\bf
J}$).


\subsection{Applied field perpendicular to the current flow}
\label{ss.anperp}

In the following, we suppose that ${\bf B}$ is almost parallel to
the HPR-LPR interface, that is very low $\varphi$ and
$\theta=\pi/2$, Fig. \ref{f.sketchGB}. For this case, the applied
field is approximately perpendicular to the current flow, so that
the vortex bows in the $z$ direction only. This is consistent with
the numerical calculations in Sec. \ref{ss.res} below.

Let define the axis $x'$ as that in the direction of the applied
field and $y'$ as that in the direction of ${\bf e}_z\times{\bf
e}_{x'}$, Fig. \ref{f.sketchGB}. Taking the coordinate $x'$ as the
vortex line parameter, its positions are
\begin{equation}
\label{zpar} {\bf r}(x')=x'{\bf e}_{x'}+z(x'){\bf e}_z,
\end{equation}
so that the $z(x')$ dependence uniquely describes the vortex line.
Using this parametrization, it can be seen from Eq. (\ref{ftau})
that the elastic force per unit length is in the ${\bf
e}_{y'}\times {\bf e}_l$ direction and, thus, it can only
compensate the driving force in this direction. Therefore, the
relevant component of the driving force is $\Phi_0[{\bf
J}\cdot{\bf e}_{y'}]{\bf e}_{y'}\times{\bf
e}_{l}=\Phi_0J\cos\varphi{\bf e}_{y'}\times{\bf e}_{l}$. The
differential equation for the vortex line of Eq. (\ref{difeq})
becomes
\begin{equation}
\label{difeqper}
J\Phi_0\cos\varphi-{\epsilon_l}{\left[{1+\left(\frac{{\rm
d}z}{{\rm d}x'}\right)^2}\right]^{-3/2}}\frac{{\rm d}^2z}{{\rm
d}x'^2}+f_p=0.
\end{equation}

\subsubsection{Infinite pinning force in the high pinning region}
\label{sss.infinitefp}

In order to simplify the system, we assume that the maximum
pinning force per unit length in the high pinning region is much
higher than any other line forces in the problem, i. e. the
pinning force in the low pinning region, the driving force and the
elastic force. With this approximation, the vortex in the HPR
follows a straight line in the ${\bf B}$ direction for any $\bf
J$. Thus, we only have to obtain the vortex line configuration in
the LPR. First, we study the case of null pinning force in the LPR
and afterwards we include the effect of a finite $f_{p,m}$.

\begin{figure}[htb]
\begin{center}
\includegraphics[width=11cm]{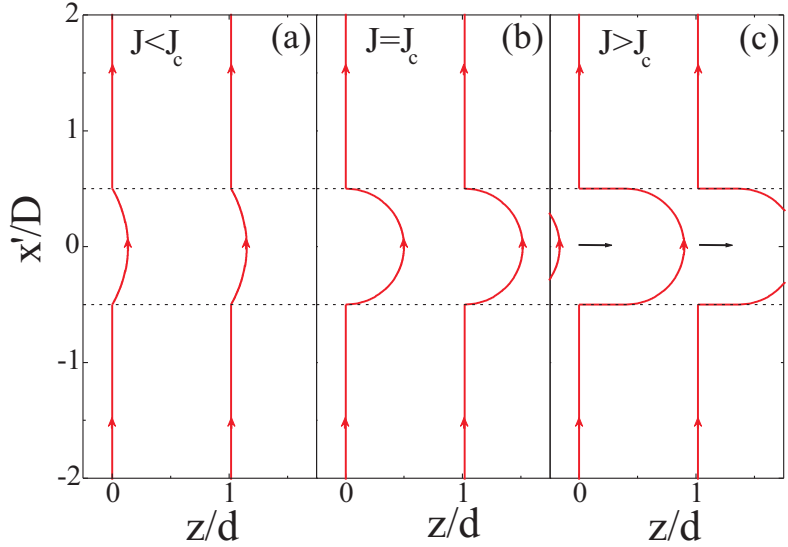}
\vspace{-0.7cm} \caption{\small Vortex bowing for the situation of
Fig. \ref{f.sketchGB} with $\theta=\pi/2$, $\varphi\ll\pi/2$ and
infinite maximum pinning force per unit length in the HPR. The
graphs correspond to $J<J_c$ (a), $J=J_c$ (b) and $J>J_c$ (c) from
left to right.} \label{f.cutper}
\end{center}
\end{figure}

Assuming null pinning force, the differential equation
(\ref{difeqper}) can be solved by direct integration. For doing
this, Eq. (\ref{difeqper}) must be first solved for $\frac{{\rm
d}z}{{\rm d}x'}(x')$ and then for $z(x')$ with a result
\begin{equation}
\label{zxper_gen} z(x')=k_1-\sqrt{R^2-(x'+k_2)^2},
\end{equation}
where $R$ is defined as $R=\epsilon_l/(|\cos\varphi|J\Phi_0)$ and
$k_1$ and $k_2$ are integration constants. These constants can be
found taking into account that the vortex line in the HPR is
fixed, obtaining that
\begin{equation}
\label{zxper} z(x')=\sqrt{R^2-(D/2)^2}-\sqrt{R^2-x'^2},
\end{equation}
where $D=d/|\sin\theta\sin\varphi|$ is the length of a straight
vortex in the LPR (Fig. \ref{f.sketchGB}). From Eq. (\ref{zxper})
we see that the vortex line bents forming a circular arch with
radius $R$. The elastic and driving forces per unit length are
calculated from Eqs. (\ref{ftau}), (\ref{fd}) and (\ref{zxper}),
yielding the result {\setlength\arraycolsep{2pt}
\begin{eqnarray}
{\bf f}_e & = & -\frac{\epsilon_l}{R}{\bf e}_{y'}\times{\bf e}_l \label{feper}\\
{\bf f}_d & = & \cos\varphi\Phi_0J{\bf e}_{y'}\times{\bf e}_l.
\label{fdper}
\end{eqnarray}}
>From these equations we see that ${\bf f}_e$ and ${\bf f}_d$ have
uniform magnitude and are perpendicular to the vortex line. The
elastic force per unit length is inversely proportional to the
curvature radius $R$. Therefore, the maximum $|{\bf f}_e|$ that
the vortex is able to produce, $f_{e,m}$, is
\begin{equation}
\label{fcut} f_{e,m}=2\epsilon_l/D
\end{equation}
that leads to a critical $J$, $J_c$, as
\begin{equation}
\label{Jcper} J_c=\frac{2\epsilon_l}{\Phi_0D\cos\varphi} =
\frac{2\epsilon_l}{\Phi_0d}\tan{|\varphi|},
\end{equation}
Where at the second equality we used that $D=d/\sin{|\varphi|}$
for $\theta=\pi/2$ (Fig. \ref{f.sketchGB}). For $J$ lower than
$J_c$ the vortex line is stable, while for $J>J_c$ it cannot
balance the driving force and the vortex line breaks, Fig.
\ref{f.cutper}. For $J>J_c$, Fig. \ref{f.cutper}(c), there is a
straight horizontal vortex segment in the border between the HPR
and LPR because in this place the driving force pushes the vortex
towards the HPR, where the available pinning force is very large
and, thus, fully compensates the driving force.

The breaking process is completed if the horizontal vortex segment
cross-joins and recombines
\cite{brandt79JLT,wagenleithner82JLT,boudiab01PRL} with the fixed
part of the neighbouring vortex. This will happen when the driving
force overcomes the repulsion between these two segments. Indeed,
not only is the repulsion between vortices at right angles strongly
suppressed but also the interaction force becomes attractive for
separations close to $\xi$ for any orientation
\cite{brandt79JLT,wagenleithner82JLT}. The expected situation after
cross-joining is that the vortex segment in the LPR joins with the
undeformed neighbouring one and there appear two horizontal
antiparallel vortex lines. These horizontal vortices will experience
a strong attraction which can overcome the pinning and driving
forces, so that the vortices move towards each other and anihilate.
In the case that the vortex attraction is not strong enough, the
elongation and recombination process repeats, creating horizontal
vortices with multiple vorticity. These vortices have a larger core
and, thus, present a lower pinning line force and their mutual
attraction is larger. At some moment these horizontal vortices will
detach from the HPR-LPR border and anihilate to each other.
Afterwards, the breaking and cross-joining process is repeated.

For large enough magnetic fields, i. e. low vortex separations,
there can be vortex cross-joining for $J$ below that of maximum
vortex deformation when the distance between the vortex arc in the
LPR and the straigh part of the next vortex is of the order of
$\xi$.

Let define the vortex breaking force $F_{\rm break}$ as the required
driving force magnitude necessary to break the vortex in absence of
pinning in the LPR. Using Eqs. (\ref{el}) and (\ref{fd}) we find
that the driving force is {\setlength\arraycolsep{2pt}
\begin{eqnarray}
\label{Fdper}
{\bf F}_{d} & = & \int_{\rm LPR}{\rm d}l{\bf f}_d =
\int_{-D/2}^{D/2}{\rm d}{x'}|\partial_{x'}{\bf r}|
\cos\varphi\Phi_0J{\bf e}_y\times\frac{\partial_{x'}{\bf r}}{|\partial_{x'}{\bf r}|} \nonumber\\
& = & \cos\varphi\Phi_0J{\bf e}_y\times\int_{-D/2}^{D/2}{\rm
d}{x'}\partial_{x'}{\bf r}=-\cos\varphi\Phi_0JD{\bf e}_z.
\end{eqnarray}}
The breaking force is obtained by taking $J=J_c$ and using Eqs.
(\ref{Jcper}) and (\ref{Fdper}), with a result
\begin{equation}
\label{Fcut} F_{\rm break}=2\epsilon_l.
\end{equation}

We next consider the presence of a pinning force in the LPR as
follows. If the driving force per unit length is lower than
$f_{p,m}$, the pinning force per unit length counteracts the driving
force and the vortex stays as a straight line. Otherwise, the vortex
bows and the elastic force per unit length compensates the remaining
force. For this situation, the vortex line follows Eq.
(\ref{difeqper}) with $f_p=-f_{p,m}$. Taking this into account, the
vortex configuration still follows Eqs. (\ref{zxper_gen}) and
(\ref{zxper}) but with a curvature radius
$R=\epsilon_l/(|\cos\varphi|J\Phi_0-f_{p,m})$. Thus, Eqs.
(\ref{feper})-(\ref{fcut}) are still valid but now $J_c$ is
\begin{equation}
\label{Jcperfp}
J_c=\frac{2\epsilon_l}{\Phi_0d}\tan{|\varphi|}+\frac{f_{p,m}}{\Phi_0|\cos\varphi|}.
\end{equation}
This equation is equivalent to that deduced by Durrell {\it et al}
in \cite{durrell03PRL,cuttingEu}, if it is taken into account that
$F_{\rm break}=2\epsilon_l$.

The vortex line of Fig. \ref{f.cutper} is nonderivable at the
border of the LPR and HPR. Then, on these points there acts a net
elastic force with magnitude $|{\bf F}_e|=\epsilon_l|{\bf
e}_{x'}-{\bf e}_l(x'=d/2)|$ [Eq. (\ref{delF})] which can only be
balanced by a finite pinning force. Such a pinning force is
possible under the assumption of infinite maximum pinning force
per unit length in the HPR, $f_{p,m}^{*}$. If $f_{p,m}^{*}$ is
limited to a finite value, the vortex line must be derivative at
every point. Thus, the vortex can bow in the HPR even if the
driving force per unit length is lower than $f_{p,m}^{*}$.
However, the main features of the vortex breaking remain the same.

We contemplate a finite pinning force per unit length in the HPR
in the following section.

\subsubsection{Finite pinning force in the high pinning region}
\label{sss.finitefp}

We next relax the assumption of infinite maximum pinning force per
unit length in the HPR, $f_{p,m}^{*}$.

The vortex configuration $z(x')$ still follows the differential
equation (\ref{difeqper}) but now $f_p$ depends on the position.
For $f_{p,m}^*>\Phi_0J>f_{p,m}$, the vortex line in the LPR bows
in the direction of the driving force and, then, $f_p=-f_{p,m}$
there. The pinning force per unit length in the HPR can, in
principle, take any value between $-f_{p,m}^*$ and $f_{p,m}^*$. If
the transport current density is reached by increasing its value
from 0 to $J$ in a quasistatic speed, the vortex line
progressively deforms from a straight line to its final stable
configuration. In this process the pinning force density opposes
with its maximum value to the driving force. Thus, in the HPR
$f_p=f_{p,m}^*$.

Taking this into account, we solve Eq. (\ref{difeqper}) using that
the vortex line must be continuous and smooth at the LPR and HPR
boundaries, with the result
\begin{equation}
\label{zxperHPR} z(x')= \left\{ \begin{array}{ll}
0 & {\rm for}\ \ x'<-x'_0 \\
\sqrt{{R^*}^2-\left( x'+x'_0 \right)^2}-R^*
& {\rm for}\ \ -x'_0<x'<-D/2 \\
-\sqrt{R^2-x'^2}+(2x'_0/D)\sqrt{R^2-D^2/4}-R^*
& {\rm for}\ \ -D/2<x'<D/2 \\
\sqrt{{R^*}^2-\left( x'-x'_0 \right)^2}-R^*
& {\rm for}\ \ D/2<x'<x'_0 \\
0 & {\rm for}\ \ x'_0<x' \\
\end{array} \right.
\end{equation}
with
{\setlength\arraycolsep{2pt}
\begin{eqnarray}
R & = & \epsilon_l/(|\cos\varphi|J\Phi_0-f_{p,m})\label{vorR}\\
R^* & = & \epsilon_l/(f_{p,m}^*-|\cos\varphi|J\Phi_0)\label{vorRs}\\
x'_0 & = & (D/2)(1+R^*/R) \label{xp0}.
\end{eqnarray}}
The quantity $x'_0$ is the minimum $|x'|$ for which the vortex is
not deformed and $R$ and $R^*$ are the curvature radii in the LPR
and the HPR, respectively.

\begin{figure}[htb]
\begin{center}
\includegraphics[width=12cm]{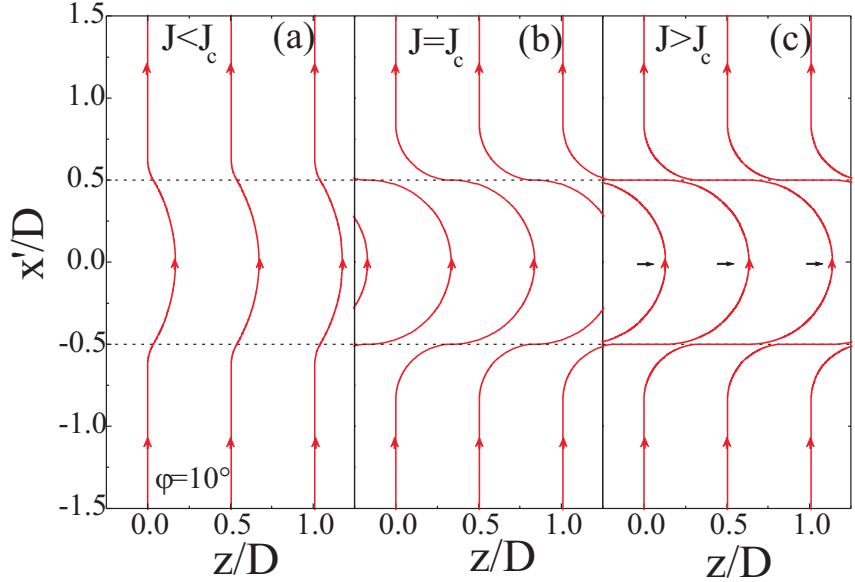}
\vspace{-0.7cm} \caption{\small Vortex bowing for the situation of
Fig. \ref{f.sketchGB} with $\theta=\pi/2$, $\varphi=10^{\rm o}$ and
a finite maximum pinning force in both the LPR and HPR $f_{p,m}$ and
$f_{p,m}^*$, respectively. The graphs are for $f_{p,m}^*/f_{p,m}=9$,
$2\epsilon_l/(df_{p,m})=18.6$, $d=10$nm and $B=6$T, corresponding to
the experimental situation in Fig. \ref{f.JcGBfit}. Half the number
of vortices have been plot for a better visualization.
 } \label{f.cutperHPR10}
\end{center}
\end{figure}

\begin{figure}[htb]
\begin{center}
\includegraphics[width=12cm]{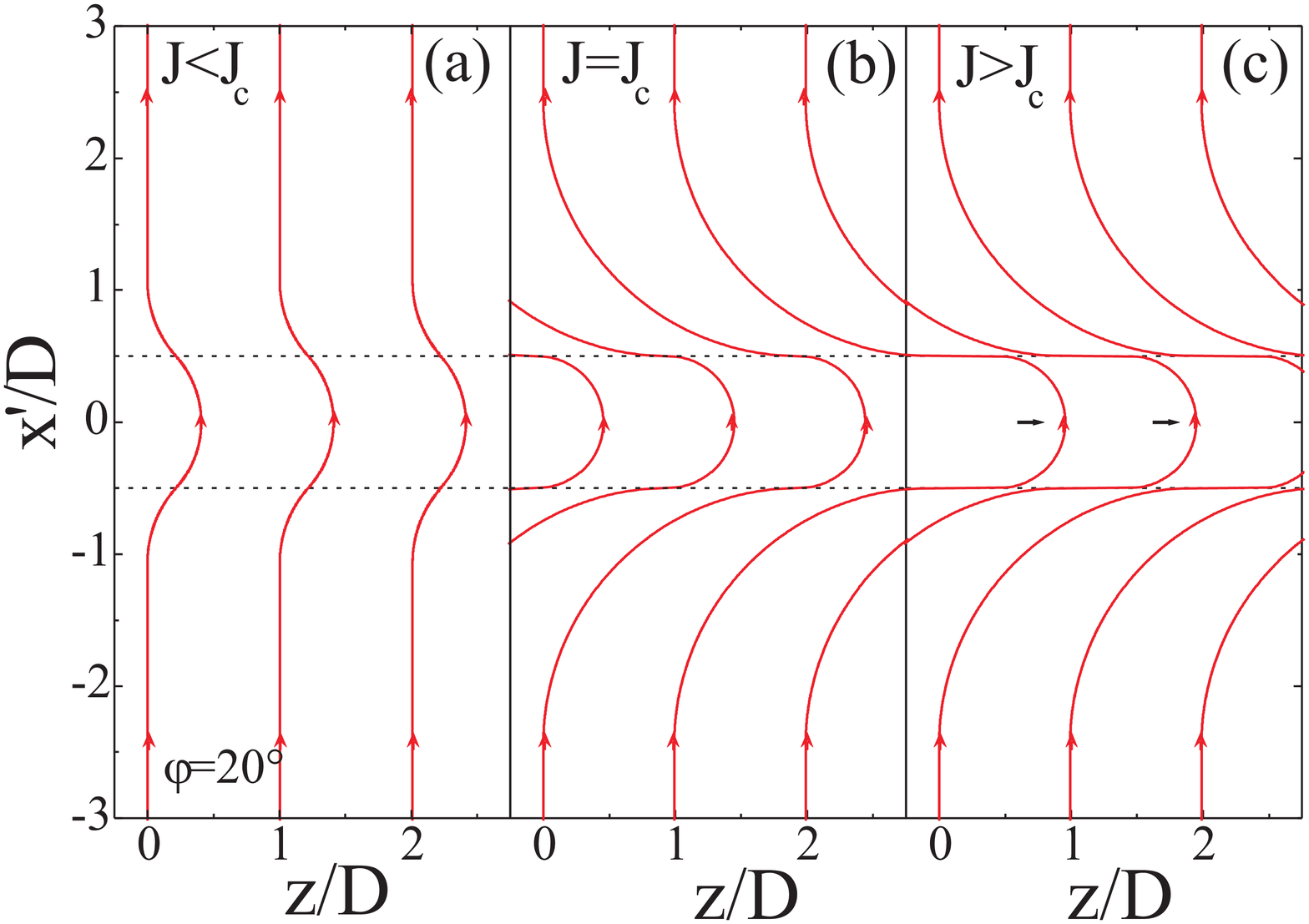}
\vspace{-0.7cm} \caption{\small The same as Fig.
\ref{f.cutperHPR10} but for $\varphi=20^{\rm o}$.}
\label{f.cutperHPR20}
\end{center}
\end{figure}

The vortex breaking process for this situation is shown in Figs.
\ref{f.cutperHPR10} and \ref{f.cutperHPR20}, where we took
$f_{p,m}^*/f_{p,m}=9$ and $2\epsilon_l/(df_{p,m})=18.6$,
corresponding to the experimental situation in Fig.
\ref{f.JcGBfit}. From Figs. \ref{f.cutperHPR10} and
\ref{f.cutperHPR20} we can see that the breaking process is
essentially the same as for infinite $f_{p,m}^*$ (Fig.
\ref{f.cutper}). However, for $J$ below that of maximum elastic
force, the vortex bending could cause a local separation between
neighboring vortices smaller than $\xi$. This would reduce $J_c$
compared with the predicted one from elasticity considerations.
For our experimental situation in Sec. \ref{ss.expGB}, the minimum
vortex distance at $J=J_c$ is around $2\xi$ so that the expected
cross-joining is still at the breaking critical current density.

For local vortex separations larger than $\xi$, such as those in
Figs. \ref{f.cutperHPR10} and \ref{f.cutperHPR20} the limiting
$J_c$ mechanism is basically the same as for infinite $f_{p,m}^*$
(Fig. \ref{f.cutper}). Thus, Eq. (\ref{fcut}) for the maximum
elastic line force is still valid. However, in order to observe
breaking for finite $f_{p,m}^*$, it is not enough that $f_d$
overcomes $f_{p,m}$ but also
\begin{equation}
f_{p,m}^*>f_{p,m}+f_{e,m}=f_{p,m}+2\epsilon_l|\sin\varphi|/d
\label{cutcond}.
\end{equation}
This can be seen as follows. The driving force per unit length is
uniform in the vortex which, from the equilibrium condition in the
LPR it is $f_d=f_{p,m}+f_{e,m}$. Then, if
$f_{p,m}^*<f_d=f_{p,m}+f_{e,m}$, there is depinning in the HPR and
there is no vortex breaking. Moreover, from equation (\ref{cutcond}) it can be seen that
$2\epsilon_l/d>f_{p,m}^*-f_{p,m}$ there will be breaking or
depinning in the HPR, depending on the value of $\varphi$. Thus,
from Eqs. (\ref{Jcperfp}) and (\ref{cutcond}), the $J_c(\varphi)$
dependence will be
\begin{equation}
J_c(\varphi)=\left\{
\begin{array}{ll}
J_c=\frac{2\epsilon_l}{\Phi_0d}\tan{|\varphi|}+\frac{f_{p,m}}{\Phi_0|\cos\varphi|}
& {\rm for}\ \varphi<\varphi_d\\
J_c=\frac{f_{p,m}^*}{\Phi_0|\cos\varphi|} & {\rm for}\
\varphi\ge\varphi_d
\end{array}
\right.
\end{equation}
with
\begin{equation}
\varphi_d=\arcsin\frac{f_{p,m}^*-f_{p,m}}{2\epsilon_l/d}.
\label{phid}
\end{equation}

\begin{figure}[htb]
\begin{center}
\includegraphics[width=12cm]{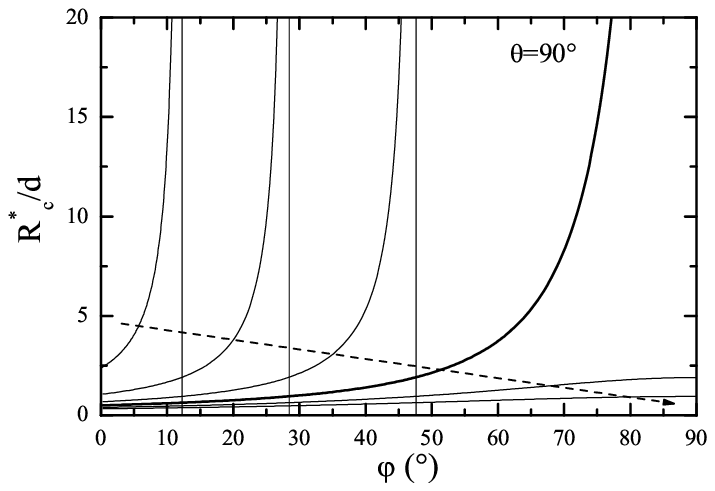}
\vspace{-0.7cm} \caption{\small Vortex line radius in the HPR for
$J=J_c$ normalized to $d$, $R^*_c/d$, as a function of $\varphi$
(Fig. \ref{f.sketchGB}) for $2\epsilon_l/(df_{p,m})=19$ and
$f_{p,m}^*/f_{p,m}=5,10,15,20,25,30$ in the arrow direction. The
line for $f_{p,m}^*/f_{p,m}=20$ diverges for $\varphi=90^{\rm o}$.}
\label{f.Rs}
\end{center}
\end{figure}

Another difference with the assumption of infinite $f_{p,m}^*$ is
that now the vortex significantly bends in the HPR, with a maximum
deformation at $J=J_c$. The curvature radius in the HPR at
$J=J_c$, $R^*_c$, can be deduced from Eqs. (\ref{vorRs}) and
(22), with a result
\begin{equation}
R^*_c/d=\frac{\epsilon_l}{f_{p,m}^*-f_{p,m}-\frac{2\epsilon_l}{d}|\sin\varphi|}.
\label{vorRsc}
\end{equation}
The value of $R^*_c/d$ is finite for the limit $\varphi=0$ and
$R^*_c/d$ approaches to infinite when $|\varphi|=\varphi_d$. The
behaviour of $R^*/d$ as a function of $\varphi$ is plotted in Fig.
\ref{f.Rs} for $2\epsilon_l/(df_{p,m})=19$ and
$f_{p,m}^*/f_{p,m}=5,10,15,20,25$ and 30 in the arrow direction. The
angle $\varphi_d$, calculated from Eq. (\ref{phid}), is
12.2,28.3,47.5 and 90$^{\rm o}$ for $f_{p,m}^*/f_{p,m}=5,10,15$ and
20, respectively. In Fig. \ref{f.Rs} it can be seen that $R^*$
diverges when it approaches to $\varphi_d$ but it has no
discontinuities for $2\epsilon_l/(df_{p,m})<f_{p,m}^*/f_{p,m}-1$.

\subsection{Applied field parallel to the current flow}
\label{ss.anpar}

We next describe the situation that the applied field is parallel
to the current flow, usually known as the force-free
configuration. For simplicity, we only study the case of infinite
pinning force per unit length in the HPR and no pinning in the
LPR.

This geometry corresponds to $\varphi=\theta=\pi/2$ in Fig.
\ref{f.sketchGB}, so that ${\bf B}=B{\bf e}_y$. Taking $y$ as the
vortex line parameter, the differential equation (\ref{difeq}) turns
into
\begin{equation}
\label{difeqpar} \frac{(\partial_y^2x{\bf e}_x+\partial_y^2z{\bf
e}_z)}{\sqrt{1+(\partial_yx)^2+(\partial_yz)^2}} +
\frac{\Phi_0J}{\epsilon_l}(\partial_yz{\bf e}_x-\partial_yx{\bf
e}_z)=0.
\end{equation}

For the force-free configuration, it has been found that the
vortices are susceptible to become spirals
\cite{walmsley72JPF,clem77PRL}. Moreover, our numerical results
below (Fig. \ref{f.vorGB} in Sec. \ref{s.num}) show that for a ${\bf
B}$ almost parallel to $\bf J$, the vortex line configuration is an
spiral with twist pitch $-d$. The spiral must cross the $y$ axis at
$y=\pm d/2$ for continuity with the vortex line in the HPR. A spiral
following these features is
\begin{eqnarray}
\label{spiralx} {\bf r}(y)=a\left( 1+\cos\frac{2\pi y}{d}
\right){\bf e}_x+y{\bf e}_y+a\sin\frac{2\pi y}{d}{\bf e}_z,
\end{eqnarray}
where $a$ is the radius of the spiral. This spiral follows Eq.
(\ref{difeqpar}) and, thus, it describes a static vortex
configuration. The spiral of Eq. (\ref{spiralx}) has the axis in
the $y$ direction but displaced a vector $a{\bf e}_x$ form it.
Actually, any rotation of the spiral from Eq. (\ref{spiralx})
around the $y$ direction is also a static vortex configuration.
The elastic and driving forces per unit length for these spiral
vortex lines are {\setlength\arraycolsep{2pt}
\begin{eqnarray}
{\bf f}_e & = & -\epsilon_l\frac{\frac{4a\pi^2}{d^2}}{1+\left(\frac{2a\pi}{d}\right)^2}{\bf e}_t \label{fepar}\\
{\bf f}_d & = &
J\Phi_0\frac{\frac{2a\pi}{d}}{\sqrt{1+\left(\frac{2a\pi}{d}\right)^2}}
{\bf e}_t \label{fdpar}
\end{eqnarray}}
with
\begin{equation}
{\bf e}_t= \cos\left(\frac{2\pi y}{d}-\phi\right){\bf
e}_x+\sin\left(\frac{2\pi y}{d}-\phi\right){\bf e}_z.
\end{equation}
The radius at which there is equilibrium $a_{\rm eq}$ is deduced
from Eqs. (\ref{fepar}) and (\ref{fdpar}) and considering ${\bf
f}_e+{\bf f}_d=0$, obtaining
\begin{equation}
\label{aeq} a_{\rm eq}=\sqrt{
\left(\frac{\epsilon_l}{J\Phi_0}\right)^2-\left(\frac{d}{2\pi}\right)^2
}.
\end{equation}
By examination of Eqs. (\ref{fepar}) and (\ref{fdpar}) we see that
the vortex at $a=a_{\rm eq}$ is in unstable equilibrium. Indeed, a
radius slightly smaller than $a_{\rm eq}$ presents an elastic
force higher than the driving force and the spiral shrinks until
it becomes an straight line. For $a>a_{\rm eq}$ the driving force
is higher than the elastic one and the vortex expands
indefinitely. However, any equilibrium is not possible when $J$ is
high enough to produce an $a_{\rm eq}$ from Eq. (\ref{aeq}) with
null or imaginary values. The minimum $J$ value at which this
occurs, $J_c$, is
\begin{equation}
\label{Jcpar} J_c=\frac{2\pi\epsilon_l}{d\Phi_0}.
\end{equation}

Comparing $J_c$ for the parallel and the perpendicular cases, Eqs.
(\ref{Jcper}) and (\ref{Jcpar}), we see that for the same vortex
length in the LPR $J_c$ for the force-free situation is only $\pi$
times higher than that one for a perpendicular field.



\section{Numerical approach}
\label{s.num}

In this section a numerical model is introduced in order to
calculate the vortex deformation in a depressed pinning layer, Fig.
\ref{f.sketchGB}, at any orientation of the macroscopic current
density $\bf J$ and the applied magnetic field ${\bf B}$. As for the
analytical approximation, we consider an elastic model for the
vortex line. In order to simplify the analysis, we assume the
approximation of infinite maximum pinning force per unit length in
the HPR, $f_{p,m}^*\to\infty$, for the numerical calculations. As
shown above, Sec. \ref{sss.finitefp}, taking into account a finite
$f_{p,m}^*$ does not substantially change the $J_c$ results.


\subsection{Vortex line calculation}

We next outline the numerical method used for calculating the vortex
line configuration. In order to generalize the procedure for several
applications, we consider a vortex segment of length
$d$ strongly pinned at its ends, beyond them the vortex line is
straight and follows the applied field direction.

The vortex configuration is calculated by physical evolution in a
similar way as in \cite{boudiab01PRL}, as follows.

We take the initial state of a straight vortex in the applied field
direction and we divide the pinning-free segment into $n$ identical
line elements. In the following discussion we term the boundary
between elements as a \emph{node}. Then, we calculate the forces per
unit length ${\bf f}_e$, ${\bf f}_d$ and ${\bf f}_p$ corresponding
to each node and change the node position proportionally to the
resulting line force. That is, the position of the node labelled as
$i$ at the time instant $k$, ${\bf r}_{i,k}$, is calculated from the
previous time instant as
\begin{equation}
\label{phev} {\bf r}_{i,k}={\bf r}_{i,k-1}+\Delta_k[{\bf f}_e({\bf
r}_{i,k-1})+{\bf f}_d({\bf r}_{i,k-1})+{\bf f}_p({\bf r}_{i,k-1})]
\end{equation}
with
\begin{equation}
\Delta_k=(t_k-t_{k-1})/\eta, \label{Deltak}
\end{equation}
where $\eta$ is a proportionality factor and $t_k$ is the time at
instant $k$. The time evolution of Eqs. (\ref{phev})-(\ref{Deltak})
has the physical basis of an asymptotic viscous vortex movement with
a viscosity per unit length $\eta$, for which $\eta\dot{{\bf
r}}={\bf f}_e+{\bf f}_d+{\bf f}_p$ \cite{boudiab01PRL,blatter94RMP}.
In the numerical method, we implemented a self-adaptative procedure
for choosing the optimum $\Delta_k$ at each time instant for a quick
convergence to the stationary situation.

The elastic and driving line forces, ${\bf f}_e$ and ${\bf f}_d$ are
calculated using Eqs. (\ref{delF}) and (\ref{fd}), respectively, and
${\bf f}_p$ is evaluated taking into account that it opposes to the
action of the driving force and has a maximum magnitude $f_{p,m}$.
Then, the effect of ${\bf f}_p$ when the driving line force is
smaller than $f_{p,m}$ is to avoid any vortex deformation for $|{\bf
f}_d|\le f_{p,m}$ and for higher ${\bf f}_{d}$, ${\bf f}_p$ reduces
the effect of the driving force by $f_{p,m}$.


\subsection{Critical current calculation}

For a vortex breaking phenomenon, the critical current $J_c$ is
the maximum $J$ for which there exists a stable vortex line
configuration. A current above $J_c$ creates a driving force per
unit length that cannot be balanced by neither the elastic nor the
pinning ones and the vortex moves or deforms indefinitely.

According to this, the critical current is calculated as follows. We
take a $J$ interval with one boundary below $J_c$, $J_{\rm min}$,
and the other one above, $J_{\rm max}$. Then, we choose a certain
current between them, $J'$, and calculate the vortex distribution.
If the vortex line for $J=J'$ converges to a stable solution, we
change $J_{\rm min}$ into $J'$ and $J_{\rm max}$ into $J'$,
otherwise. Thus, the interval containing $J_c$ is narrowed. In order
to ensure a solution of the procedure, we choose a very wide initial
interval, with $J_{\rm min}=10^3$ A/m$^2$ and $J_{\rm max}=10^{15}$
A/m$^2$. We found that the number of vortex line calculations is
minimized by taking $J'$ as the middle point in logarithmic scale,
$J'=\sqrt{J_{\rm min}J_{\rm max}}$. The number of vortex line
evaluations for obtaining a 1\% wide interval is around 12.

Numerical calculations showed that the $J_c$ result do not
significantly change with the number of vortex elements $n$. The
computing time for $n=20$ in an standard table computer is of few
seconds (around 15s), allowing a systematic study of the $J_c$
dependence with the applied magnetic field orientation.

\subsection{Results}
\label{ss.res}

\subsubsection{Vortex bending}
\label{sss.resvor}

Although the numerical method described above is valid for any
relative orientation between $\bf J$ and ${\bf B}$, here we restrict
our study for ${\bf B}$ in the $xy$ plane of Fig. \ref{f.sketchGB},
that is, $\theta=\pi/2$. This corresponds to the experimental
situation of a thin film YBCO bicrystal with the applied field
parallel to the surface and, consequently, the $ab$ planes.

\begin{figure}[hbp]
\begin{center}
\includegraphics[width=9cm]{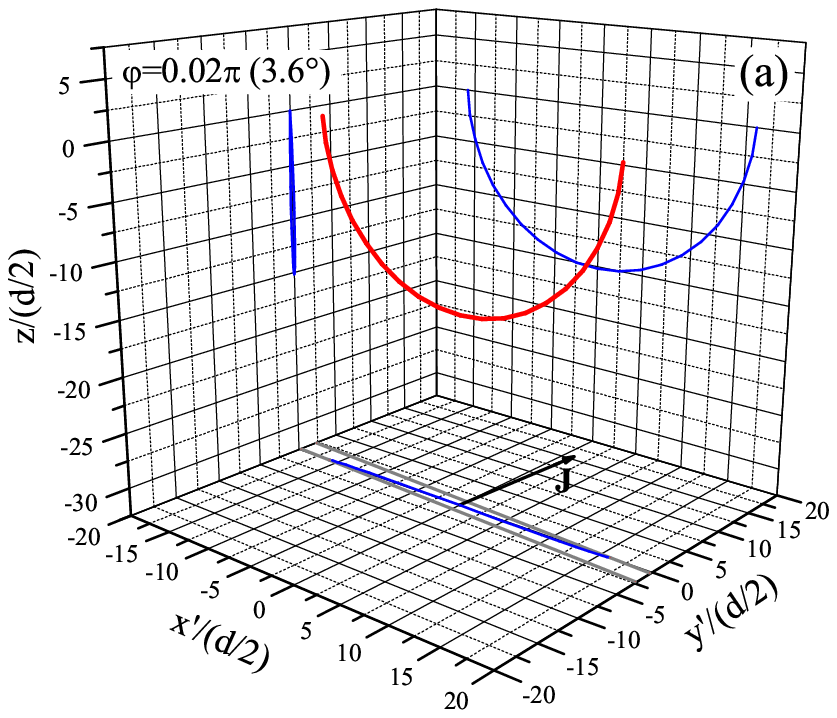}
\includegraphics[width=9cm]{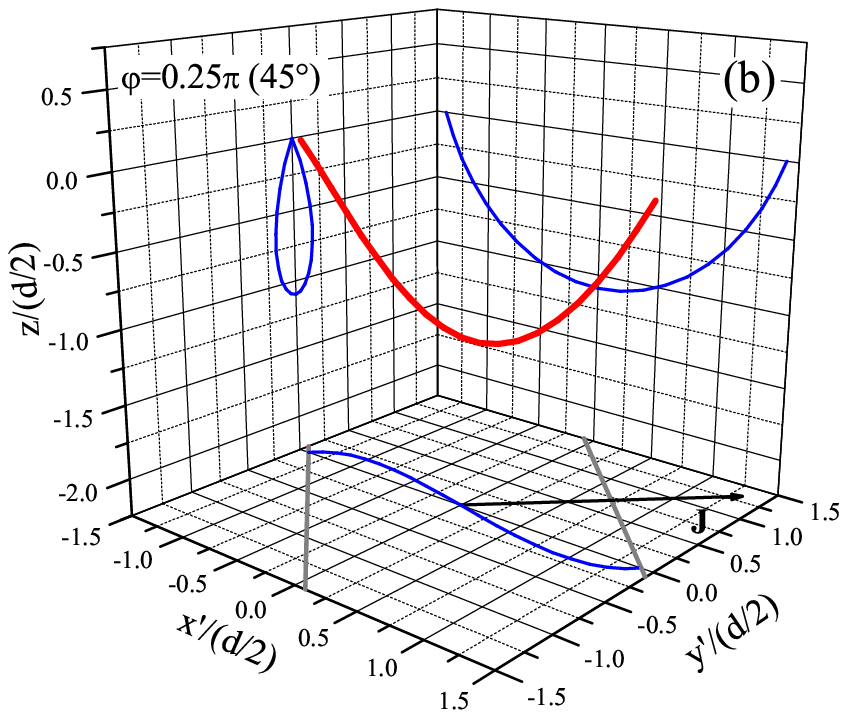}
\includegraphics[width=9cm]{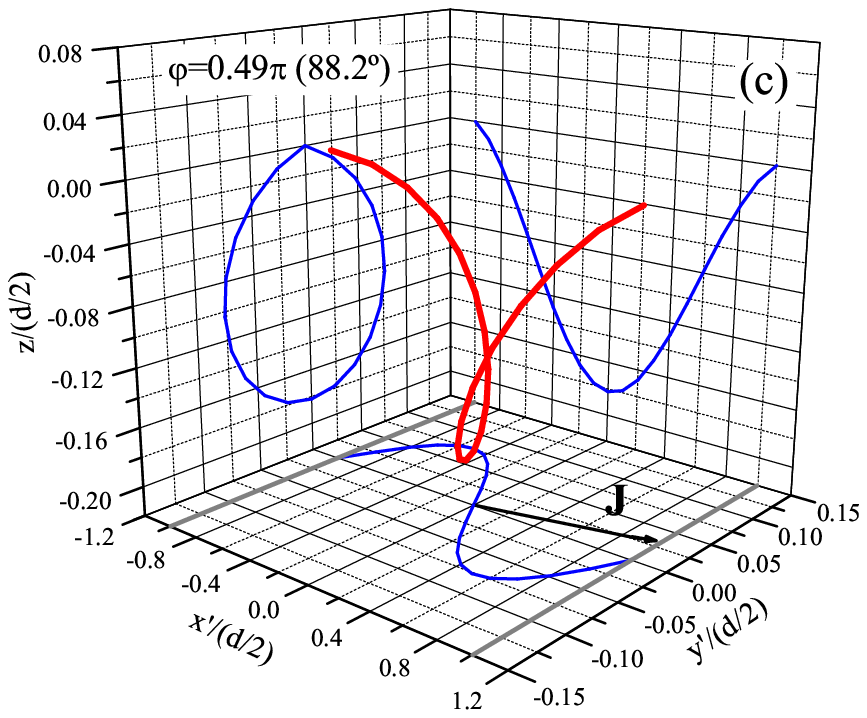}
\vspace{0cm} \caption{\small Vortex deformation for the situation of
Fig. \ref{f.sketchGB} at $J=J_c$ for $\theta=\pi/2$ and several
$\varphi$ angles (red thick lines), where all dimensions have been
normalized to the GB half width $d/2$. Blue thin lines are the
projections in the $x'z$, $x'y'$ and $y'z$ planes, the black arrow
shows the current direction and the grey lines show the LPR-HPR
boundaries in the $x'y'$ plane. In (c), note the different scales of
the axis.} \label{f.vorGB}
\end{center}
\end{figure}

In Fig. \ref{f.vorGB} we present the vortex bending at $J=J_c$ for
several $\varphi$ orientations. In the plots, the $x'$ axis is in
the direction of the applied magnetic field and the $y'$ axis is
perpendicular to both $x'$ and $z$, Fig. \ref{f.sketchGB}.

The situation in Fig. \ref{f.vorGB}(a) is for $\bf J$ almost
perpendicular to the field, with $\varphi=0.02\pi$$(3.6^{\rm o})$.
It can be seen that the vortex line bends almost exclusively in
the $z$ direction, following a half circle with a diameter equal
to the vortex length in the LPR, $D$, as predicted analytically in
Sec. \ref{ss.anperp}. When $\bf J$ makes an intermediate angle
with ${\bf B}$, as in Fig. \ref{f.vorGB}(b), the vortex also bends
antisymmetrically in the $y'$ direction and the maximum
deformation in the $z$ axis is smaller than $D$. Another
interesting situation is when $\bf J$ is almost parallel to ${\bf
B}$, as shown in Fig. \ref{f.vorGB}(c) for which $\varphi=0.49\pi
(88.2^{\rm o})$. For this limit, the vortex approaches to an helix
with twist pitch $-D$. This result justifies the helical vortex
assumption for the analytical deduction in Sec. \ref{ss.anpar}.

\subsubsection{Critical current density}

In the following we study first $J_c$ with null pinning in the LPR
and then we consider the effect of a finite $f_{p,m}$. Below, we
refer to $J_c$ for $f_{p,m}=0$ as the breaking current density,
$J_{\rm break}$.

\begin{figure}[tbp]
\begin{center}
\includegraphics[width=11cm]{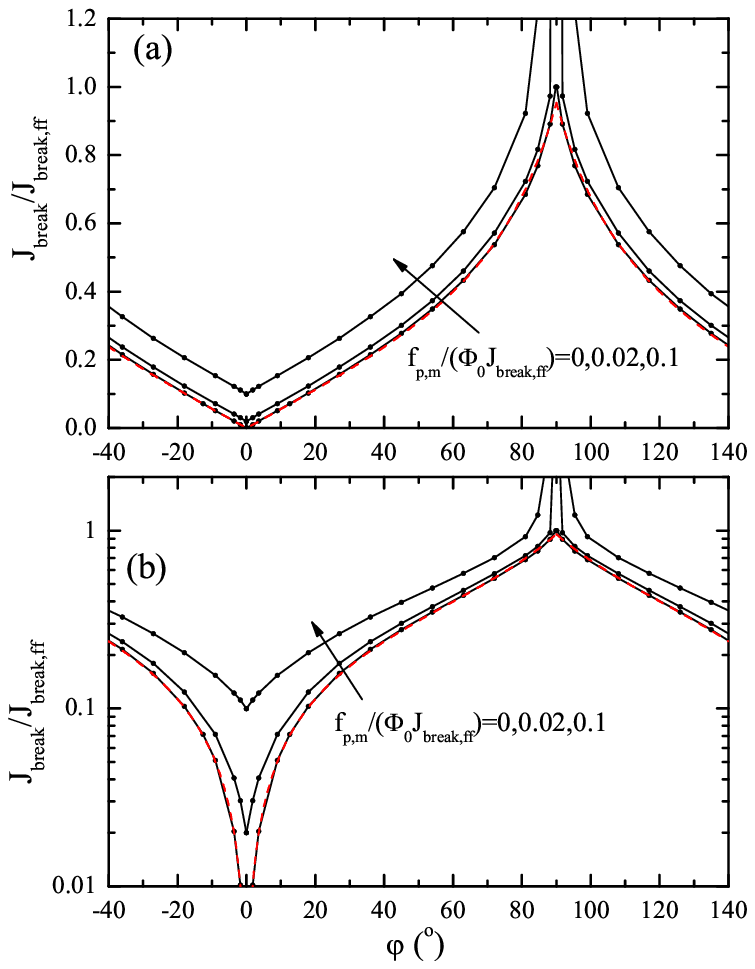}
\vspace{-0.5cm} \caption{\small Calculated critical current
density, $J_c$, for $\theta=\pi/2$ as a function of $\varphi$ for
the situation of Fig. \ref{f.sketchGB}. The curves are universal,
for any $d$ and $\epsilon_l$ when, $J_c$ is normalized to the
critical current density in the force-free configuration $J_{\rm
{break, ff}}$, expressed by Eq. (\ref{Jcpar}). Lines plus symbols
are for numerically calculated data for $f_{p,m}/(\Phi_0J_{\rm
break,ff})=$0,0.02,0.1 from bottom to top and the dash red line is
for the analytical fit of Eq. (\ref{Jcutfit}). } \label{f.JcGBnum}
\end{center}
\end{figure}

In order to study the general breaking behaviour in a GB, in Fig.
\ref{f.JcGBnum} (solid line plus symbols) we present $J_{\rm
break}$ normalized to the breaking current density at the
force-free situation $J_{{\rm break,ff}}=2\pi\epsilon_l/(d\Phi_0)$ [Eq. (\ref{Jcpar})] as a function
of the angle $\varphi$ (see sketch in Fig. \ref{f.sketchGB}). Such
representation is universal for any $d$ and $\epsilon_l$. This can
be seen as follows. $J_{\rm break}$ is the maximum $J$ at which
the elastic line force compensates the driving one. Since the
elastic line force is proportional to $\epsilon_l$, the same must
be for $J_{\rm break}$. Besides, the whole vortex line
configuration at $J=J_c$ scales with the LPR thickness $d$. Since
$f_e$ is proportional to the vortex line curvature and twisting,
$f_e$ and, thus, $J_{\rm break}$ must be inversely proportional to
$d$. Consequently, $J_{\rm break}$ is proportional to
$\epsilon_l/d$ and to $J_{\rm break,
ff}=2\pi\epsilon_l/(d\Phi_0)$.

The maximum $J_{\rm break}$ (Fig. \ref{f.JcGBnum}) is for
$\varphi=\pi/2$, corresponding to $\bf J$ parallel to ${\bf B}$.
As can be seen in the figure, the $J_{\rm break}$ value at the
maximum is $J_{\rm break,ff}$, in accordance to the analytical
deduction in Sec. \ref{ss.anperp}.

The numerically calculated $J_{\rm break}/J_{\rm break,ff}$ of
Fig. \ref{f.JcGBnum} can be well fitted by the expression
\begin{equation}
 J_{\rm break}/J_{c,ff}=\frac{1}{\pi}\tan{|0.78\varphi|}+\frac{0.22}{\pi}\sin{|\varphi|},
\label{Jcutfit}
\end{equation}
corresponding to the dash line in Fig. \ref{f.JcGBnum}. The
difference between the analytical fit and the numerical data
decreases with decreasing $\varphi$, being of a $5\%$ for
$\varphi=\pi/2$ and less than $1.3\%$ for $|\varphi|\le 0.4\pi
(72^{\rm o})$.

We next study the effect of including a pinning force in the LPR. In
Fig. \ref{f.JcGBnum} we plot $J_c/J_{c,ff}$ for a nonzero $f_{p,m}$
[$f_{p,m}/(\Phi_0J_{c,ff})=0.02,0.1$], showing a roughly uniform
increase of $J_c$ by $f_{p,m}/\Phi_0$ in most of the $J_c(\varphi)$
curve except close to $\varphi=\pi/2$, where $J_c$ diverges. This
behaviour can be explained as follows. If $f_{p,m}$ is large enough
to counteract the driving force, the vortex stays as a straight
line. Otherwise, the vortex bends and the elastic force adds to the
pinning one. However, the deformation of the vortex does not always
lead to a compensation of the driving force. Indeed, for $\bf J$
almost parallel to ${\bf B}$ and high enough $J$, the vortex deforms
helically and experiences a higher driving force than when it was
straight. Then, for this situation the critical current density is
determined by depinning, so that $J_c=f_{p,m}/(\Phi_0|\cos
\varphi|)$ and $J_c$ diverges for $\varphi=\pi/2$. For the
orientations that flux breaking exists, the increase in the
numerically calculated $J_c$ due to pinning in the LPR fits well to
$f_{p,m}/(\Phi_0\sqrt{|\cos\varphi|})$. Thus, taking into account
the above features and Eq. (\ref{Jcutfit}) for $J_{\rm break}$,
$J_c$ can be well described by
\begin{eqnarray}
J_c = {\rm max}(J_{c,1},J_{c,2}) \label{Jcwpin}
\end{eqnarray}
with
\begin{eqnarray}
J_{c,1} & = & \frac{f_{p,m}}{\Phi_0|\cos\varphi|} \label{Jc1}\\
J_{c,2} & = & \frac{2\epsilon_l}{d\Phi_0}\left(\tan{|0.78\varphi|}
+ 0.22\sin{|\varphi|}\right) +
\frac{f_{p,m}}{\Phi_0\sqrt{|\cos\varphi|}}. \label{Jc2}
\end{eqnarray}
Comparing Eq. (\ref{Jc2}) and equation (\ref{Jcperfp}) for low
magnetic field orientations $\varphi$, it can be seen that for the
low-$\varphi$ limit they are coincident and they differ less than
2\% and 8\% for $\varphi<15^{\rm o}$ and  $\varphi<30^{\rm o}$,
respectively.

If we take into account that the pinning line force in the HPR has
a finite value, the critical current density cannot be larger than
that limited by depinning in the HPR (Sec. \ref{sss.finitefp}).
Thus, $J_c$ will have the lower value between those given by Eqs.
(\ref{Jcwpin})-(\ref{Jc2}) and $f_{p,m}^*/(\Phi_0|\cos\varphi|)$.


\section{Application to experimental situations}
\label{s.exp}

In this section we apply the theoretical model developed in Secs.
\ref{s.analit} and \ref{s.num} to the experimental situations of
LAGB and vicinal films in YBCO. The deduced $J_c(\theta,\varphi)$
is fitted to the measured data in order to obtain the vortex line
tension in these systems.

\subsection{Low-angle grain boundaries}
\label{ss.expGB}

The study of low-angle grain boundaries is of great practical
importance because they are present in YBCO coated conductors
\cite{evetts04SST}, which are the most promising high-temperature
superconducting materials for applications in electrical devices
\cite{hull03RPP}. Indeed, grain boundaries are limiting the
critical current density in coated conductors.

The geometry described in this article can be directly applied in
order to model low-angle grain boundaries in YBCO bicrystals. In
these grain boundaries there are Abrikosov-Josephson vortices
(AJV) \cite{gurevich02PRL}, which magnetic extent and normal core
is of larger size than the usual Abrikosov votices and, thus, the
AJV in the grain boundary presents lower pinning than the AV in
the grains \cite{gurevich02PRL,diaz98PRL}. Assuming a sharp
transition between the grain and the grain boundary, this
situation corresponds to the HPR/LPR/HPR layered structure in Fig.
\ref{f.sketchGB}. Actually, for the Abrikosov-Josephson vortices
in LAGB introduced in Ref. \cite{gurevich02PRL} it is assumed that
the vortices in the crystals are conventional AV and parallel to
the boundary plane. The AJV theory in Ref. \cite{gurevich02PRL}
is, then, directly applicable to high-temperature superconductors
for magnetic inductions and LAGB parallel to the $c$
crystallographic axis. For $\bf B$ in the $ab$ plane, the vortices
in the crystals are not AV but Josephson string vortices (SV),
which complicates the description near a LAGB. However, it is
expected that SV in a LAGB will present a larger phase core and
magnetic extent in the LAGB direction. This significantly changes
the SV properties, as follows.

The supercurrents around SVs flow in two characteristic
penetration depths, $\lambda_{ab}$ in the $c$ axis direction and
$\lambda_c=\Gamma\lambda_{ab}$ in the $ab$ plane
\cite{blatter94RMP,brandt95RPP}. In SVs, the phase core have a
size of $d_{\rm il}$ in the $c$ axis and $\Gamma d_{\rm il}$ in
the $ab$ planes, where $d_{\rm il}$ is the separation between $ab$
layers. SV are, then, highly anisotropic. A LAGB perpendicular to 
the SV and parallel to the $c$ axis will enlarge both the penetration depth and
the phase core in the $c$ direction, reducing the SV anisotropy.
Furthermore, SVs present a strong intrinsic pinning towards
driving forces in the $c$ direction owing to the reduced order
parameter between $ab$ planes. Therefore, increasing the SV phase
core in the $c$ direction due to a LAGB will significantly reduce
the intrinsic pinning. If the SVs are making a certain angle
$\varphi$ with the LAGB, the situation is even more complex but
for not very large $\varphi$ it is expected a significant
intrinsic pinning reduction with a weak $\varphi$ dependence.

\subsubsection{Comparison with experiments}

\begin{figure}[htp]
\begin{center}
\includegraphics[width=12cm]{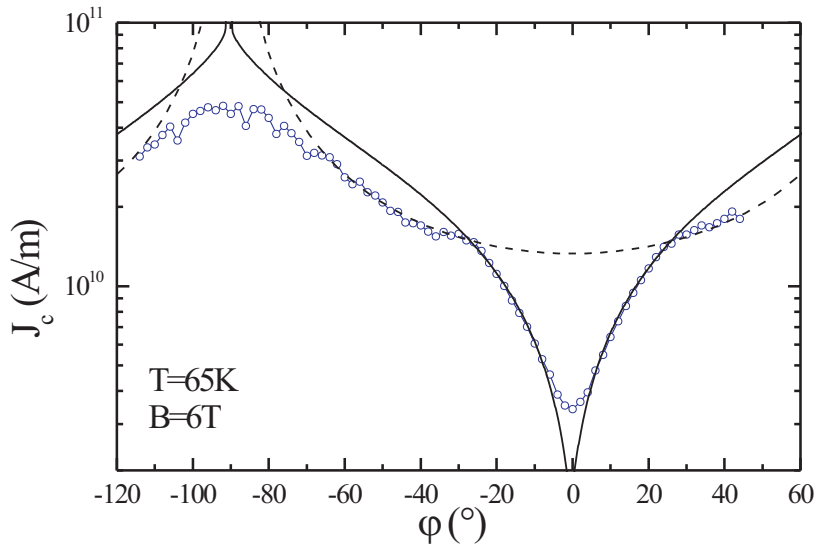}
\vspace{-0.5cm} \caption{\small Measured $J_c$ for a grain
boundary in an YBCO bicrystal with a $4.9^{\rm o}$ misorientation
angle \cite{durrell03PRL} (hollow blue circles) together with the
theoretical fit for breaking in the LAGB (solid line), calculated
from Eqs. (\ref{Jcwpin})-(\ref{Jc2}), and for depinning in the
crystals (dash line), from $J_c=f_{p,m}^*/(\Phi_0|\cos\varphi|)$.
The fit parameters are $\epsilon_l=2.89\times 10^{-13}{\rm N}$,
$f_{p,m}=3.1\times 10^{-6}{\rm N/m}$ and $f_{p,m}^*=2.8\times
10^{-5}{\rm N/m}$.} \label{f.JcGBfit}
\end{center}
\end{figure}

The $J_c$ numerical results from Fig. \ref{f.JcGBnum} (Sec.
\ref{ss.res}), or the equivalent formulae
(\ref{Jcwpin})-(\ref{Jc2}), can be used for fitting experimental
data for grain boundaries and extract the vortex line tension
$\epsilon_l$. In Fig. \ref{f.JcGBfit} we fit the experimental
$J_c(\varphi)$ dependence for a low angle GB in an YBCO bicrystal
\cite{durrell03PRL} with our numerical curve, obtaining
$\epsilon_l=F_{\rm break}/2=2.89\times 10^{-13}{\rm N}$,
$f_{p,m}=3.1\times 10^{-6}{\rm N/m}$ and $f_{p,m}^*=2.8\times
10^{-5}{\rm N/m}$ for a temperature $T=65{\rm K}$ and an applied
magnetic field $B=6{\rm T}$. From the fitting parameters and Eq.
(\ref{phid}) a maximum flux breaking angle of $\varphi_d=25.5^{\rm
o}$ is obtained, corresponding to the measured value in Fig.
\ref{f.JcGBfit}. Moreover, the experimental results from
\cite{durrell03PRL} are field independent in the measured range (1
to 5T). This can be explained by the low vortex separation compared
to $\lambda_{ab}$, which causes that the line tension is mainly the
vortex core energy, as discussed in Sec. \ref{ss.epl}.

It is interesting to compare the fitted value of $\epsilon_l$ with
the expected one for an anisotropic vortex. Using a $\lambda_{ab}$
temperature dependence of
\begin{equation}
\lambda_{ab}=\lambda_{ab,0}(1-T/T_c)^{-1/2} \label{lamT}
\end{equation}
 with
$\lambda_{ab,0}=1.50\times 10^{-7}$m and $T_c=91$K, Eq.
(\ref{elSVt}) yields $\epsilon_l\approx 1.72\times 10^{-11}$N,
which is two orders of magnitude larger than the value obtained
from the experimental data. However, a better agreement is found
when assuming an isotropic vortex with a magnetic size
$\lambda_{c}=\Gamma\lambda_{ab}$, obtaining $\epsilon_l\approx
\epsilon_{l,0}=1.38\times 10^{-13}$N. This is consistent with a
significant vortex anisotropy loss due to a vortex size increase
in the $c$ direction, caused by the LAGB (Sec. \ref{ss.expGB}).
However, the origin of this discrepancy might be the
oversimplified analysis in the deduction of Eqs.
(\ref{elSVt})-(\ref{elSVp}) for the line tension. A more accurate
prediction would require a further complicated discussion on the
vortex lattice elasticity
\cite{sudbo91PRL,sudbo91PRB,brandt95RPP}.

\subsection{YBCO films with miscut substrate}
\label{ss.expvic}

Another system which spontaneously presents flux breaking is an
YBCO film on a miscut substrate \cite{durrell04PRB,cuttingEu},
commonly named a vicinal film. It is well known that for applied
fields with an orientation angle with the $ab$ planes below around
$11^{\rm o}$, the vortex line in YBCO have a stair-like shape, composed
by pancake vortices (PV) in the
$ab$ planes and string vortices (SV) between these planes
\cite{blatter94RMP}. While pancake vortices are strongly pinned,
this is not the case for the string ones. Although a large pinning
force is required to depin SVs from the $ab$ plane direction, they
are almost free of pinning for forces parallel to the $ab$ planes.
The result is that the SV is a weakly pinned vortex segment
confined between two $ab$ planes with its ends strongly pinned by
pancake vortices. Therefore, under the presence of a transport
current with a component in the $c$ direction, the SVs will bend
in the $ab$ plane and, eventually, break for a transport current
density above a certain threshold\footnote{If the YBCO film is
grown on a normal substrate, the transport current flows parallel
to the $ab$ planes and there is no breaking phenomenon. For this
reason, it is required a miscut angle in the substrate in order to
obtain a certain component in the $c$ direction and observe vortex
breaking.}, $J_c$.

\begin{figure}[htb]
\begin{center}
\includegraphics[width=7cm]{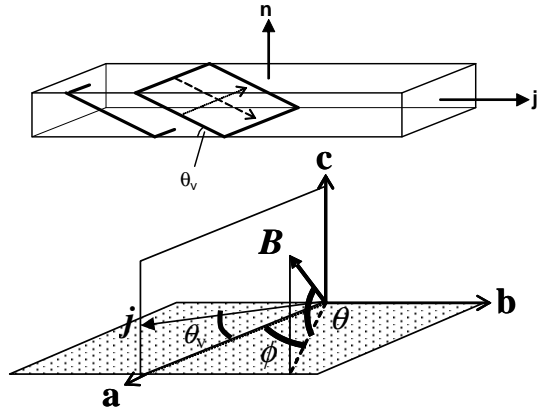} \vspace{-0.2cm}
\caption{\small Sketch of a vicinal film and description of the
angular coordinates $\theta$ and $\varphi$.} \label{f.skvic}
\end{center}
\end{figure}

\subsubsection{Breaking description for vicinal films}

\begin{figure}[htb]
\begin{center}
\includegraphics[width=16cm]{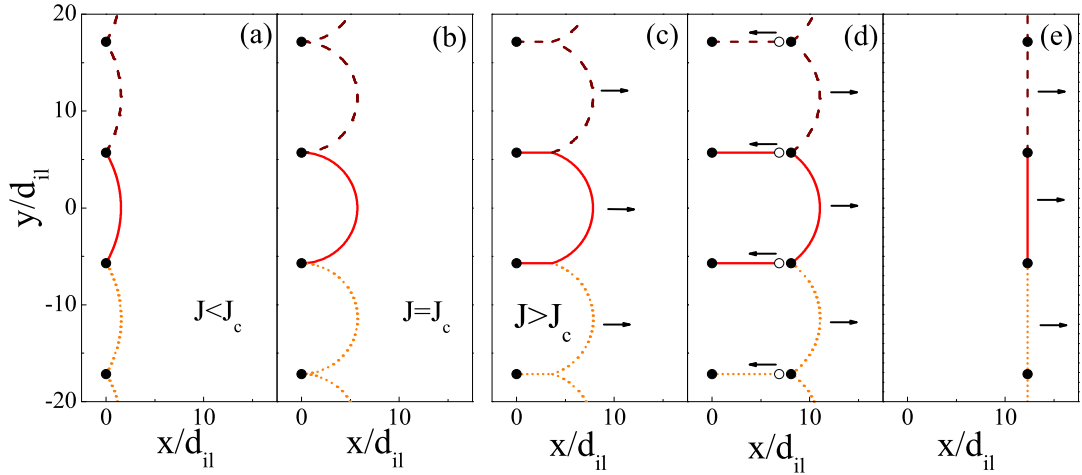} \vspace{-0.5cm}
\vspace{-0.5cm} \caption{\small Static vortex deformation (a,b)
and proposed breaking and cross-joining process (c,d,e) for a
string vortex (SV) in a vicinal film with a field orientation
$\theta=10^{\rm o}$, Fig. \ref{f.skvic}. The graph is the
projection in the $ab$ plane of a kinked vortex, where lines are
SV between $a$-$b$ planes and circles are pancake vortices. The
solid (red) lines are for SVs in the graph plane, dash (maroon)
ones are for SV at lower planes and dot (orange) lines are SV in
higher planes. Solid and hollow dots are pancake vortices and
antivortices, respectively. } \label{f.vorvic}
\end{center}
\end{figure}

Due to intrinsic pinning, the driving force will deform the SV
mainly in the $ab$ plane following a circular arch, as shown in
Fig. \ref{f.vorvic}(a,b). If we restrict our analysis to the
vortex deformation in the $ab$ planes only, a SV would expand
until it finds another string vortex in the same plane. However,
flux crossing will occur much sooner if we consider the
interaction between SV in different planes. For $J=J_c$, Fig.
\ref{f.vorvic}(b), the SVs between different planes are
antiparallel at the junction with the common pancake vortex, where
they experience a strong attractive interaction. This attraction
compensates, at least partially, the bowing effect of the driving
force and antiparallel straight SV segments appear
\ref{f.vorvic}(c). When the length of these segments is large
enough, the SV attraction will overcome the intrinsic pinning and
they will cross-join, creating a pair of unpinned
pancake-antipancake vortices \ref{f.vorvic}(d). The new pancake
follows the whole vortex drift, while the antivortex approaches to
the original pinned pancake, anihilating to each other. Finally,
we obtain a depinned vortex with undeformed SV segments, Fig.
\ref{f.vorvic}(e), for a driving force much lower than that one
required for pancake depinning.

This vortex breaking and cross-joining process is set up when the
driving line force is larger than the sum of the maximum elastic
and pinning line forces. Using the angular coordinates of Fig.
\ref{f.skvic}, the driving line force in a SV is ${\bf
f}_d=\Phi_0J\sin\theta_v{\bf b}$, where $\theta_v$ is the vicinal
angle and $\bf b$ is the unit vector in the $b$ axis. From Eq.
(\ref{fcut}), we obtain that the maximum elastic line force is
${\bf f}_e=-(2\epsilon_l/d_{\rm il})|\tan\theta|{\bf b}$, where
$d_{\rm il}$ is the $a$-$b$ interlayer distance. Considering this
and a certain pinning force, we obtain the following critical
current density
\begin{equation}
\label{Jcvic}
J_c=\frac{2\epsilon_l}{d\Phi_0\sin\theta_v}|\tan\theta|+\frac{f_{p,m}}{\Phi_0\sin\theta_v},
\end{equation}
consistent with the expression in Refs.
\cite{cuttingEu,durrell04PRB} with a total SV breaking force of
$2\epsilon_l$.

\subsubsection{Comparison with experiments}

\begin{figure}[htp]
\begin{center}
\includegraphics[width=12cm]{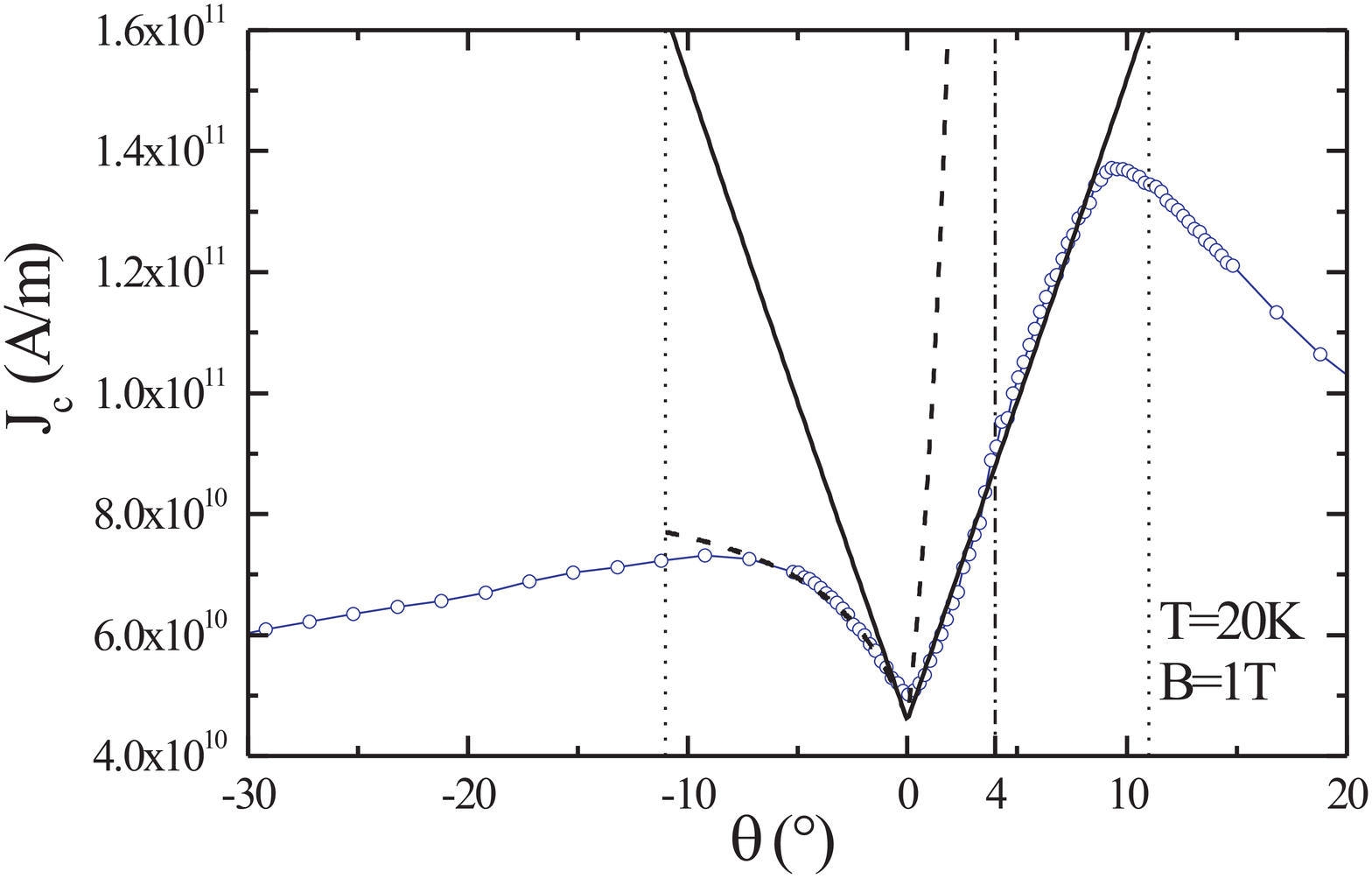}
\vspace{-0.5cm} \caption{\small Measured $J_c(\theta)$ for an YBCO
vicinal film with $4^{\rm o}$ miscut substrate \cite{durrell04PRB}
(line with circles) together with the fitting curve for the
breaking limited $J_c$ of Eq. (\ref{Jcvic}) (solid line) and the
fit for a kinked vortex depinnig (dash line), calculated using Eq.
(\ref{Jcvicpin}). The fitting parameters are
$\epsilon_l=1.69\times 10^{-14}$N, $f_{p,{\rm str}}=6.62\times
10^{-6}$N/m and $f_{p,{\rm pc}}=1.82\times 10^{-4}$N/m.}
\label{f.Jcvicfit}
\end{center}
\end{figure}

The model introduced above can be applied in order to obtain
$\epsilon_l$ for a SV, as follows. In Fig. \ref{f.Jcvicfit} we
present the measured $J_c(\theta)$ for an YBCO vicinal film with
$4^{\rm o}$ miscut substrate \cite{durrell04PRB}. In this figure we
include the fitting curve for the breaking limited $J_c(\theta)$ of
Eq. (\ref{Jcvic}) (solid line) together with the fitted
$J_c(\theta)$ for a kinked vortex depinnig (dash line), calculated
using \cite{durrell04PRB}
\begin{equation}
\label{Jcvicpin}
J_c=\frac{f_{p,{\rm str}}\cos\theta+f_{p,{\rm
pc}}\sin|\theta|}{\Phi_0\sin|\theta-\theta_v|},
\end{equation}
where $f_{p,{\rm str}}$ and $f_{p,{\rm pc}}$ are the pinning force
per unit length for the string and pancake vortices, respectively,
and $\theta_v$ is the substrate miscut angle. The measured data of
Fig. \ref{f.Jcvicfit} is optimally fitted by Eqs. (\ref{Jcvic})
and (\ref{Jcvicpin}) with $\epsilon_l=1.69\times 10^{-14}$N,
$f_{p,{\rm str}}=6.62\times 10^{-6}$N/m and $f_{p,{\rm
pc}}=1.82\times 10^{-4}$N/m. The two vertical dot lines in Fig.
\ref{f.Jcvicfit} delimit the expected angular region where kinked
vortices can exist ($\theta\approx 11^{\rm
o}$)\cite{blatter94RMP}, while the dash-dot line indicates the
$\theta$ value at which the kinked vortex presents null total
driving force, $\theta=\theta_v$, so that the depinning-limited
$J_c$ diverges. Thus, it is clear that for this situation the
limiting mechanism of $J_c$ must be flux breaking. Besides, both
Eqs. (\ref{Jcvic}) and (\ref{Jcvicpin}) for string vortex breaking
and kinked vortex depinning, respectively, fit to the experimental
data for the range $|\theta|<9^{\rm o}$; a slightly smaller value
than the originally expected $|\theta|\approx 11^{\rm o}$.

The expected SV line tension in the parallel direction from Eqs.
(\ref{elSVp}) and (\ref{lamT}) is $\epsilon_l\approx 1.9\times
10^{-12}$N, which is around two orders of magnitude larger than the
obtained one from Fig. \ref{f.Jcvicfit}. This discrepancy, also
found in Sec. \ref{ss.expGB}, evidences that the use of the
simplified Eqs. (\ref{elc})-(\ref{elSVp}) is probably too na\"\i ve
and a more detailed study of the line tension in terms of the flux
line lattice elasticity is required for a fine prediction.


\section{Conclusions}
\label{s.concl}

One way of studying the superconducting vortex lattice properties is
by vortex breaking experiments in nonuniform pinning force
distributions.

In this article, the vortex bending and breaking in a
high-pinning/low-pinning/high-pinning layered structure has been
studied from a microscopical point of view in order to extract
vortex internal properties from breaking experiments. Our study
has been based on an elastic model assuming a constant vortex line
tension, $\epsilon_l$. For the general 3D bending, this model is
applicable for isotropic superconductors in the single-vortex
approximation. The latter approximation can be done for a magnetic
penetration depth $\lambda$ much larger than the vortex separation
because for this case the vortex line energy is mainly the core
energy. As a consequence, the line tension is independent on both
the vortex deformation amplitude and the deformation length.
Within the assumption of constant line tension, the model is
applicable to any nonuniform pinning force or current density
distribution.

Summarizing, the main results from the elastic model are the
following.

When the pinning force in the high-pinning regions (HPRs) is much
higher than that in the low-pinning ones (LPR), the vortex only
bends in the LPR. For low magnetic field orientation angles with
the HPR-LPR boundary ($\varphi$), the vortex takes the shape of a
circular arch with a minimum curvature radius of half the LPR
thickness. The latter situation corresponds to the maximum elastic
force. For a transport current $J$ higher than the value which
this happens ($J_c$), the elastic force cannot balance the driving
one and the vortex moves until it finds its neighbour. After this,
we have proposed that the adjacent vortices cross-join and
recombine with an steady average vortex movement in the LPR. For
higher $\bf B$ orientation angles $|\varphi|$, there is a similar
behaviour, with the difference that the vortex deforms in two
ortogonal directions. When $\varphi$ is close to $\pi/2$,
corresponding to the force-free configuration, the vortex
experiences a helical deformation with a small radius. For the
case that $\varphi$ is exactly $\pi/2$, the vortex presents a
helical instability for $J$ above a certain threshold and keeps as
a straight line for $J$ below.

The more realistic case of finite pinning force in the HPR
presents vortex bending in both in the LPR and the HPRs. For this
case, if the maximum elastic line force in the LPR is larger than
the pinning line force difference, the vortices present depinning
in the HPR above a certain orientation angle $\varphi_d$. When the
orientation angle is slightly lower than $\varphi_d$, the vortex
bends in the HPR over a thickness larger than that of the LPR.

The model calculations have been used for obtaining the vortex
line tension in two experimental situations: a low-angle grain
boundary (LAGB) in an YBCO bicrystal and an YBCO film on a miscut
substrate. In these systems the vortices involved are vortices of
Abrikosov-Josephson nature and string vortices.

For LAGB with a low angle $\varphi$ between the magnetic field and
the grain boundary plane, the vortex mainly bends in one single
direction and the constant line tension assumption is reasonable.
The line tension obtained by fitting the modelled curve with the
measured data is much smaller than the expected one for an
anisotropic vortex from Eq. (\ref{elSVt}), although $\epsilon_l$ is
of the same order of magnitude as an isotropic vortex with
penetration depth $\lambda_c$. This result is consistent with an
anisotropy loss due to a vortex enlargement in the $c$
crystallographic axis. However, it must be kept in mind that the
applicability of the simplified equations
(\ref{elSVt})-(\ref{elSVp}) is not evident for the studied LAGB
situation.

The results for YBCO films with a miscut angle show again that the
line tension is two orders of magnitude lower than expected. This
reinforces the hypothesis that Eqs. (\ref{elSVt})-(\ref{elSVp}) are
not applicable for SVs at the studied temperature and magnetic field
conditions, for which the vortices are strongly overlapped. Thus, a
more complex analysis in terms of the flux line lattice elasticity
should be done for performed for a proper prediction of the line
tension.

In conclusion, modelling based on the vortex elastic properties
provides a good explanation for the flux breaking phenomena. In
this article, the elastic theory has been successfully applied for
obtaining the line tension of ``exotic" vortices, such as string
vortices between $ab$ planes and in low-angle grain boundaries.


\end{document}